\documentclass[pdflatex,sn-vancouver-num]{sn-jnl}

\usepackage{graphicx}%
\usepackage{multirow}%
\usepackage{amsmath,amssymb,amsfonts}%
\usepackage{amsthm}%
\usepackage{mathrsfs}%
\usepackage[title]{appendix}%
\usepackage{xcolor}%
\usepackage{textcomp}%
\usepackage{manyfoot}%
\usepackage{booktabs}%
\usepackage{algorithm}%
\usepackage{algorithmicx}%
\usepackage{algpseudocode}%
\usepackage{listings}%
\usepackage{threeparttable}
\usepackage{multirow}
\usepackage{wrapfig}

\theoremstyle{thmstyleone}%
\theoremstyle{thmstyletwo}%

\theoremstyle{thmstylethree}%

\begin{document}

\title[Statistical methods for extrapolation from one population to another: a review]{Statistical methods for {\color{black}extrapolation from one population to another}: a review}


\author*[1]{\fnm{Lingjiao} \sur{Wang}}\email{Lingjiao.Wang.1@warwick.ac.uk}

\author[2]{\fnm{Fran\c{c}ois} \sur{Mercier}}

\author[3a]{\fnm{Astrid} \sur{Jullion}}

\author[3b]{\fnm{Xuan} \sur{Zhu}}

\author[4]{\fnm{Christian} \sur{R{\"o}ver}}

\author[5,6]{\fnm{Marc} \sur{Vandemeulebroecke}}

\author[7]{\fnm{Martin} \sur{Posch}}

\author[8]{\fnm{Sarah} \sur{Zohar}}

\author*[1]{\fnm{Nigel} \sur{Stallard}}\email{N.Stallard@warwick.ac.uk}

\affil*[1]{\orgdiv{Warwick Clinical Trials Unit, Warwick Medical School}, \orgname{University of Warwick}, \orgaddress{\city{Coventry}, \country{UK}}}

\affil[2]{\orgname{Roche}, \orgaddress{\city{Basel}, \country{Switzerland}}}

\affil[3a]{\orgname{Novartis Pharma AG}, \orgaddress{\city{Basel}, \country{Switzerland}}}

\affil[3b]{\orgname{Novartis Pharmaceuticals Corporation}, \orgaddress{\city{East Hanover}, \country{USA}}}

\affil[4]{\orgname{University Medical Center G{\"o}ttingen}, \orgaddress{\city{G{\"o}ttingen}, \country{Germany}}}

\affil[5]{\orgname{Radboud university medical centre}, \orgaddress{\city{Nijmegen}, \country{Netherlands}}}

\affil[6]{\orgname{Bayer BCC AG}, \orgaddress{\city{Basel}, \country{Switzerland}}}

\affil[7]{\orgname{Medical University of Vienna}, \orgaddress{\city{Vienna}, \country{Austria}}}

\affil[8]{\orgname{INSERM}, \orgaddress{\city{Paris}, \country{France}}}


\abstract{\color{black}
	{\it Background}: Extrapolation refers to statistical approaches that support inference on treatment effects in a target population using data from completed clinical trials conducted in a source population. This is particularly important when only limited data are available in the target population. Extrapolation may be based on multiple sources of information, including pharmacokinetic, pharmacodynamic, mechanistic, or clinical outcome data. This review focuses on methods that enable borrowing of information from one population to support inference in another. 
	
	{\it Methods}: We conducted a review of statistical methodologies for extrapolation, focusing on approaches that incorporate outcome data from source populations to support inference in target populations. Methods were identified across a range of settings, including extrapolation across age groups and across different clinical indications. Literature searches were conducted using the Science Citation Index Expanded database within Web of Science and included all English-language records available up to 30 June 2025. 
	
	{\it Results}: The review identified 218 extrapolation approaches from 159 papers, which can be broadly classified into three main categories: Bayesian, frequentist, and causal inference methods. These methods vary in their assumptions, implementation, and applicability across different extrapolation settings. 
	
	
	{\it Conclusions}: This review provides a comprehensive and up-to-date overview of statistical methods for extrapolation, offering a foundation for further methodological development and applied research. It also discusses key considerations for method selection and current limitations, and highlights priorities for future research, including the development of new extrapolation methods and clinical trial designs that facilitate extrapolation.

	}




\maketitle

\section{Introduction}
\label{intro}

Up to 36 million people in the European Union are affected by one of more than 7,000 distinct rare diseases, approximately 80\% of which are of genetic origin, with 70\% of these starting in childhood \cite{ECPublicHealthRare}. The evaluation of new medicines for rare diseases, including paediatric rare diseases, is challenging. The small size of patient populations limits the feasibility of conducting adequately powered randomised clinical trials, makes it difficult to obtain sufficient evidence on safety and efficacy, and complicates the assessment of the overall benefit-risk balance.

In addition, the development of treatments in these settings is constrained by significant scientific, methodological, and ethical challenges \cite{institute2010transforming}. There is an ethical requirement to minimise both the number of clinical studies conducted in children and the number of children recruited to studies \cite{roth2011ethical,caldwell2004clinical}.  Furthermore, the number of assessments performed in paediatric studies should be limited, particularly when these investigations impose a substantial burden on patients, resulting in fewer longitudinal efficacy measurements being available in paediatric populations.  Similar constraints arise in other rare disease and small-population settings.
While data on drug safety and efficacy may be available from larger patient populations, or from similar drugs, important differences between populations exist in drug exposure and/or pharmacodynamic effects. 

One approach to addressing these challenges is the extrapolation of treatment effects, pharmacokinetic/pharmacodynamic (PK/PD) data, and real-world data within the drug development process \cite{ICH_E11A}.  Extrapolation allows information and conclusions derived from studies in one or more subgroups of a patient population (the {\it source} population) to be used to make inferences in another subgroup (the {\it target} population) \cite{european2013concept}. This strategy, often referred to as {\it borrowing information} or {\it bridging}, can reduce the required sample size of studies in target populations, decrease the need for additional clinical studies, and accelerate patient access to new medicines \cite{RN1390,fda2016pediatric,ema2018reflection}.  

Under the guidelines issued by the International Council for Harmonisation of Technical Requirements for Pharmaceuticals for Human Use (ICH), extrapolation requires sufficient justification that the source and target populations are relevant to one another with respect to factors affecting disease progression and treatment response \cite{ICH_E11A,ICH_E11R}. {\color{black}The nature and extent of this justification may vary according to the extrapolation setting. For example, when extrapolating across age groups, similarity may be assessed through developmental and pharmacological considerations, whereas extrapolation across indications may rely on evidence relating to disease mechanisms, treatment effects, and other clinical or biological factors. When sufficient similarity or relevance can be established, extrapolation may be used to support evidence on the safety and effectiveness of medicinal products in the target population.} 
{\color{black} A similar concept arises in causal inference, where extrapolation may be viewed as a problem of generalisability (from a study population to a broader population) or transportability (from a source population to a distinct target population). Both settings require assumptions regarding the relationship between the source and target populations, including consistency, exchangeability, and positivity\cite{Hernan2020}. Positivity requires sufficient overlap between the source and target populations in relevant covariates. Violations of this assumption may limit the validity of extrapolation. 

In practice, however, extrapolation is often considered in settings where similarity between populations is incomplete or uncertain, rather than only when similarity has been fully established. Consequently, a wide range of statistical extrapolation methods have been developed to borrow information from source populations while accounting for uncertainty and potential differences between populations.}

The purpose of this paper is to review methods for extrapolation from one population to another, with a particular focus on supporting the inference of treatment effects in the target population. Extrapolation is considered both across age groups and across indications, for example from adults to children, from non-rare to rare diseases, or from larger to smaller patient populations. Although several reviews have discussed related extrapolation methods\cite{wadsworth2018extrapolation,gamalo2017statistical}, an updated methodological review is warranted to summarise recent developments and to provide an overview of the current landscape of statistical approaches for extrapolation, as this field is rapidly evolving. 
We summarise the literature by categorising papers into broad methodological themes and critically assessing how these methods can be applied, thereby providing a resource for researchers and practitioners facing the challenges associated with extrapolation. This review also provides a foundation for future research and for the development of new extrapolation methods. 
{\color{black}In this review, the terms `{\it extrapolation}' and `{\it borrowing}' are used interchangeably to refer to the use of information from a source population to support inference in a target population.}

A number of extrapolation strategies lie outside the primary focus of this review. These include predicting long-term effects from short-term outcomes within the same population\cite{fleming1996surrogate,RN490}, assessing the consistency of treatment effects in multi-regional clinical trials\cite{guideline1998ethnic,RN2012}, and applying PK/PD modelling to predict dosing and efficacy in target populations\cite{RN1390,RN461,ICH_E11A,ema2018reflection}. 
In particular, within these excluded PK/PD-based methods, Bayesian feedback, commonly used in model-informed precision dosing to update individual-level parameters\cite{vozeh1987estimates,svensson2019individualised,maier2021reinforcement}, is not considered.

\section{Review methodology}
\label{methods}

Literature searches were conducted using the Science Citation Index Expanded (SCI-EXPANDED) database within the Web of Science (\href{https://www.webofscience.com/wos/alldb/smart-search}{https://www.webofscience.com}). {\color{black}
Given the methodological focus of this review, SCI-EXPANDED was selected because it provides broad coverage of high-quality, peer-reviewed statistical, methodological, and biomedical journals.}
Searches included all records available up to 30 June 2025 and were restricted to English-language records. Separate searches were performed to identify potentially relevant papers proposing statistical methods for: (a) incorporating historical data into {\color{black}current} clinical trials; (b) using data on short-term endpoints from one population to support inference on long-term outcomes in another population; (c) paediatric and bridging clinical trials. 
Search terms, which were applied to the title, abstract, and keywords of candidate papers, are given in Appendix \ref{search terms}. 

{\color{black}Records identified through the literature searches were screened in two stages. First, titles and abstracts were reviewed to identify records that were potentially eligible for inclusion. This initial screening was conducted by the first author. Second, full texts of potentially eligible papers were assessed against the inclusion and exclusion criteria. Full-text screening and eligibility assessment were conducted by one of the first and last authors, with any uncertainties resolved through discussion and consensus. During the full-text review stage, additional relevant papers were identified through citation searching of included papers and assessed for eligibility using the same criteria.}

\begin{table}[h]
	\centering
	\caption{\label{exclusion criteria}Inclusion and exclusion criteria for paper selection.}
		\renewcommand{\arraystretch}{1}
		\footnotesize
		\setlength{\tabcolsep}{2mm}
			\begin{tabular}{p{0.45\linewidth}|p{0.45\linewidth}}
				\toprule
				\textbf{Inclusion criteria} & \textbf{Exclusion criteria}\\
				\midrule
				(1) Published in English. & (1) Indirect treatment comparisons or network meta-analyses. \\
				(2) Indexed in the Web of Science up to 30 June 2025.  & (2) No primary outcomes reported for the target population in current study. \\ 
				(3) Applied statistical methods to borrow information across different populations. & (3) Assessed only consistency between populations.  
				\\
				(4) Developed or implemented new statistical methods for data analysis, rather than prediction. & (4) Borrowed information solely from different endpoints within the same population.  \\
				(5) Focused on methods for extrapolation, rather than trial design.  &  (5) Used control arms from the source population directly as those of the target population. \\ 
				(6) Included phase II or later clinical trials evaluating treatment efficacy. & (6) Employed only traditional Bayesian or simple frequentist pooling methods.  \\
				\bottomrule
		\end{tabular}
\end{table}

Inclusion and exclusion criteria are defined in Table \ref{exclusion criteria}. 
These were chosen to focus the review on statistical methods that enable extrapolation to support inference in the target population within a confirmatory trial setting, 
leading to the exclusion of papers addressing phase I or dose-finding problems. 
We also excluded papers that did not have primary outcomes in the target population and papers that assessed only consistency between populations.
In addition, we excluded papers that simply incorporated control data from the source population, using either pooling or standard Bayesian methods, without accounting for potential differences between populations. 
Papers presenting no new methods, such as methodological comparison studies or papers that described applications of previously proposed methods, were excluded.

From each included paper, we extracted details of all statistical methods relevant to extrapolation.  
When identical methods were described in multiple papers, the earliest was retained.

\section{Summary of review results}
\label{review summary}

\begin{figure}[h]
	\centering
	\includegraphics[width=9cm]{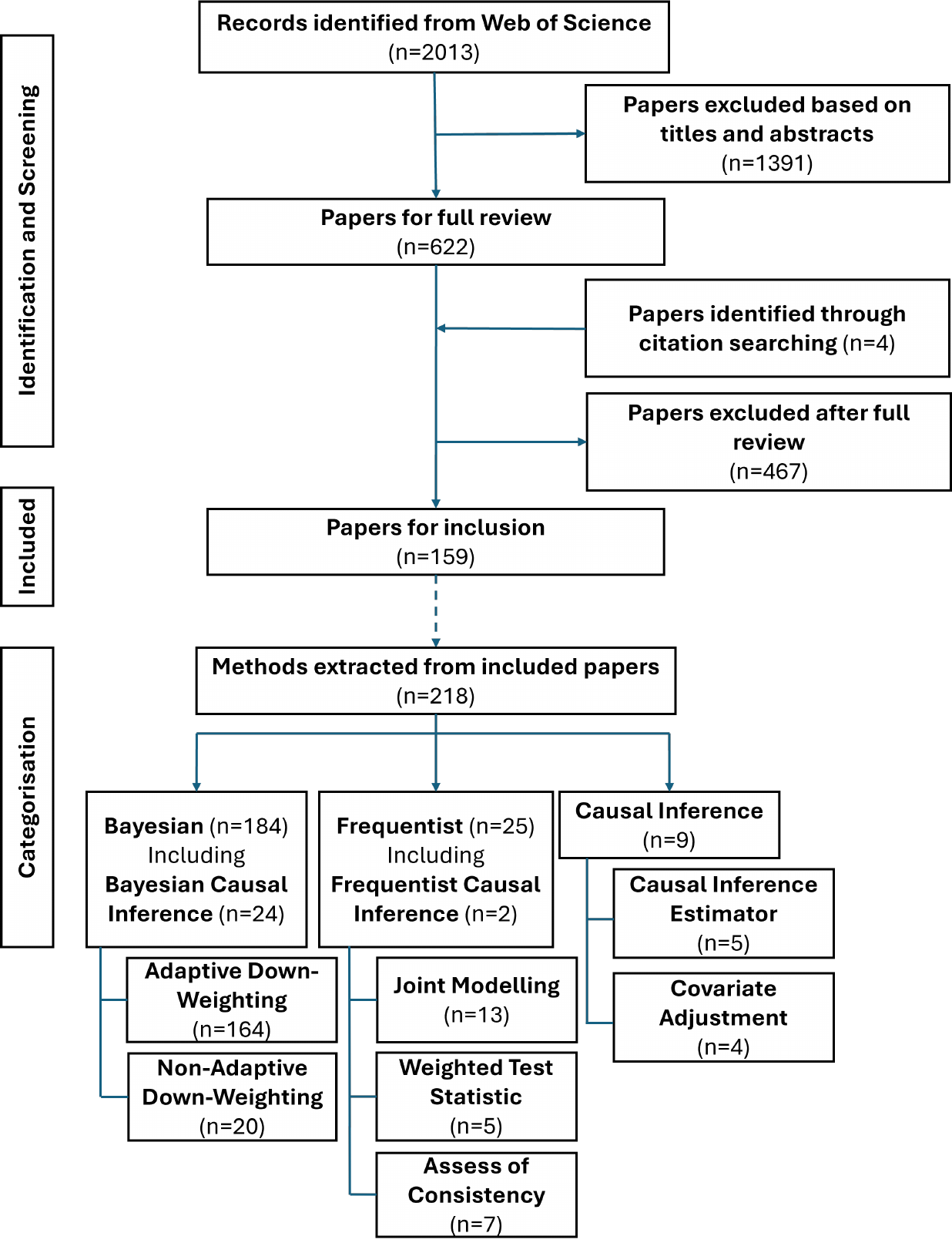}
	\caption{PRISMA diagram of selection and categorisation of papers in the review of methods.}
	\label{PRISMA_flow_diagram}
\end{figure}

{\color{black}The study screening and selection process is summarised in the PRISMA flow diagram\cite{page2021prisma} presented in Figure \ref{PRISMA_flow_diagram}. The figure also summarises the categorisation of the included papers and the methods identified in the review.} Following the removal of duplicates, our searches identified 2013 papers. After applying the inclusion and exclusion criteria, 159 papers were retained. Of these, four papers, labelled `Papers identified through citation searching' in Figure \ref{PRISMA_flow_diagram}, were identified through citation searching of included papers rather than through the initial database searches. A total of 218 distinct statistical methods extracted. These methods were categorised into three main areas: (i) Bayesian borrowing methods (184/218); (ii) frequentist approaches (25/218); and (iii) causal inference methods (9/218).  

{\color{black}Although the distinction between Bayesian and frequentist approaches is not always straightforward in  applications, Bayesian methods generally combine prior distributions with observed data through posterior updating, whereas frequentist methods rely on sampling distributions and likelihood-based inference without specifying prior distributions\cite{Ruberg2021,PeckCampbell2019}. Frequentist pharmacometric analyses may also incorporate simulation-based uncertainty quantification, confidence intervals, prediction intervals, and bootstrap procedures\cite{Kummel2018}. 
These inferential principles were used to classify the methods identified in this review. Methods that explicitly addressed extrapolation through the frameworks of generalisability or transportability and employed covariate adjustment techniques were classified as causal inference approaches.
} Many methods combine causal inference techniques with Bayesian or frequentist borrowing approaches by first adjusting source population data for baseline covariates and then applying the borrowing framework. In this review, such methods are referred to as Bayesian causal inference methods or frequentist causal inference methods and are classified within the Bayesian or frequentist categories, respectively.  {\color{black}Although these categories are not mutually exclusive, this classification provides a practical framework for organising the extrapolation literature reviewed here.} 

These categories were further refined into seven broad groups. Bayesian methods comprised: (1) constructing informative prior distributions for parameters of the target population by adaptively down-weighting the source population data, and (2) constructing informative prior distributions for parameters of the target population by non-adaptively down-weighting such data. Here, adaptive down-weighting refers to incorporating external data with weights that vary according to the commensurability {\color{black}(i.e., the degree of similarity or compatibility)} between target and source populations, whereas non-adaptive down-weighting refers to the use of fixed, pre-specified weights.
Frequentist methods comprised: (1) jointly modelling data from source and target populations; (2) combining data across populations using weighted test statistics; and (3) assessing the commensurability of parameters between the populations before borrowing.  
Causal inference methods comprised: (1) combining data across populations using causal inference estimators, and (2) incorporating data from both populations through covariate adjustment. 
Further details of the extrapolation methods are provided in Section \ref{methods identified}, including a brief overview of each methodological area and key references. 

\begin{figure}[h]
	\centering
	\includegraphics[width=12cm]{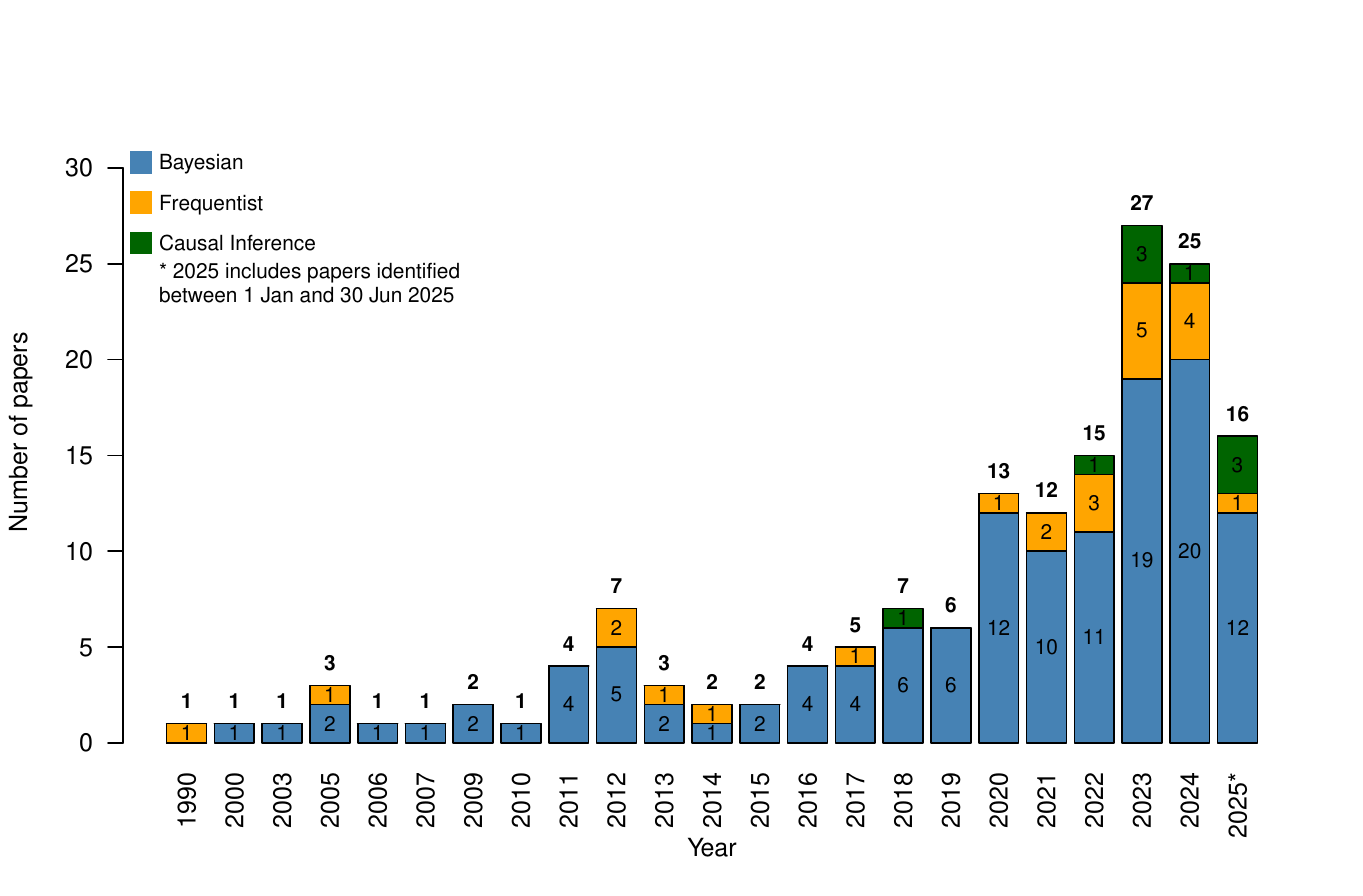}
	\caption{Number of published papers by year and methodological category: Bayesian, frequentist, causal inference, with papers for 2025 identified between 1 January and 30 June only.}
	\label{5_year_diagram}
\end{figure}

{\color{black}Figure \ref{5_year_diagram} illustrates the number of papers published each year between 1990 and 2025, with papers for 2025 identified between 1 January and 30 June only. 
No papers were published between 1991 and 1999. Research activity was limited between 2000 and 2010 but increased steadily thereafter, with a marked rise in recent years. The growth in Bayesian methods is particularly pronounced, with more than 10 papers published annually between 2020 and 2025. Specifically, 19 papers were published in 2023 and 20 in 2024. Between 1 January 2025 and 30 June 2025, a further 12 papers were identified. 
Although the number of papers describing frequentist and causal inference approaches remains lower than that for Bayesian methods, both areas have shown a steady increase in research activity in recent years, particularly between 2022 and 2025. 

As the overall volume of scientific papers has increased over time, these trends should be interpreted as descriptive summaries of the identified literature rather than as evidence of changes in the relative popularity of different methodological approaches.}

\section{Thematic analysis of methods for extrapolation}
\label{methods identified}

In this section, a discussion of the identified extrapolation methods is presented. 
For each extrapolation method, the key concepts are first introduced, followed by a discussion of papers that proposed, extended, or refined the method. 

We first introduce the notation used throughout the summary. In all method descriptions, parameters and data from the source (target) population are indexed by the subscript $S$ ($T$). The parameter of interest, such as a treatment effect, is denoted by $\theta_S$ ($\theta_T$) for the source (target) population. Accordingly, $x_S$ ($x_T$) represents data from a source (target) population, which depends on $\theta_S$ ($\theta_T$). When $\theta_S$ and $\theta_T$ are assumed to be equal, their common value is denoted by $\theta$. The sample size for observations from a source (target) population is denoted by $n_S$ ($n_T$).
When multiple datasets are available from source populations, let $H$ denote the total number of source datasets. Then, $x_{Sh}$ and $n_{Sh}$ represent the data and the sample size of the $h$th dataset, respectively, for $h = 1, \dots, H$.


\subsection{Bayesian methods}

The searches described above identified 184 Bayesian methods from 127 papers\cite{RN430,RN316,RN162,RN167,RN397,RN333,RN418,RN310,RN790,RN665,RN64,Baumann03072025,RN409,RN146,RN23,RN170,RN784,RN104,RN196,RN338,RN470,RN918,RN328,RN311,RN325,RN82,RN61,RN91,RN359,RN357,RN309,RN330,RN340,RN218,wang2019propensity,baron2022bayesian,RN644,RN6,RN83,RN158,RN52,RN742,RN203,RN459,RN455,RN363,RN754,RN181,RN549,RN139,RN94,RN217,RN18,RN807,RN432,RN159,RN150,RN599,RN33,RN600,RN554,RN225,RN448,RN40,RN49,RN661,RN995,RN344,RN103,RN909,RN326,RN318,RN123,RN73,RN673,RN361,RN922,RN887,RN343,RN1162,RN8,RN9,RN933,RN16,RN508,RN882,RN481,RN535,RN472,RN594,RN908,RN552,RN81,RN912,RN883,RN362,RN374,hobbs2018bayesian,RN910,RN539,bean2023bayesian,RN498,RN437,RN207,RN484,RN315,RN141,RN160,RN562,RN551,RN442,RN1097,RN352,RN317,RN306,RN237,RN914,RN369,RN715,RN1419,RN499,RN138,RN428,RN816,RN651,RN2011,RN168}. 
Among these, 164 methods\cite{RN430,RN316,RN162,RN167,RN397,RN333,RN418,RN310,RN790,RN665,RN64,Baumann03072025,RN409,RN146,RN23,RN170,RN784,RN104,RN196,RN338,RN470,RN918,RN328,RN311,RN325,RN82,RN61,RN91,RN359,RN357,RN309,RN330,RN340,RN218,wang2019propensity,baron2022bayesian,RN644,RN6,RN83,RN158,RN52,RN742,RN203,RN459,RN455,RN363,RN754,RN181,RN549,RN139,RN94,RN217,RN18,RN807,RN432,RN159,RN150,RN599,RN33,RN600,RN554,RN225,RN448,RN40,RN49,RN661,RN995,RN344,RN103,RN909,RN326,RN318,RN123,RN73,RN673,RN361,RN922,RN887,RN343,RN1162,RN8,RN9,RN933,RN16,RN508,RN882,RN481,RN535,RN472,RN594,RN908,RN552,RN81,RN912,RN883,RN362,RN374,hobbs2018bayesian,RN910,RN539,bean2023bayesian,RN498,RN437,RN207,RN484,RN315,RN141,RN160,RN562,RN551,RN442,RN1097,RN352,RN317,RN306,RN237,RN914} accounted for potential differences between $\theta_T$ and $\theta_S$ by formulating priors for $\theta_T$ that, when updated with emerging data from the new trial, adaptively weighted $x_S$ according to the commensurability between $x_S$ and $x_T$. The remaining 20 methods\cite{RN311,RN430,RN369,RN715,RN1419,RN499,RN138,RN428,RN816,RN651,RN2011,RN168} adopted a fixed (non-adaptive) approach to down-weighting $x_S$. 

All methods in this category augment data from a new trial in the target population ($x_T$) with existing data from the source population ($x_S$).  
In this context, differences between $\theta_T$ and $\theta_S$ may arise due to variations in trial protocols, advances in medical care, or demographic differences between the populations. More generally, source population data are considered informative about $\theta_T$ only when the clinical effects of treatment in different populations are sufficiently similar, although this similarity is typically uncertain and must be justified or assessed within the chosen methodological framework. 
{\color{black}Let $\pi(\theta)$ denote the prior distribution for $\theta$, and $\pi(\theta|x)$ denote the corresponding posterior distribution given the observed data $x$. 
In many Bayesian borrowing approaches, the prior for the target population is constructed using source data $x_S$. Although, $\pi(\theta|x_S)$ is formally a posterior distribution conditional on $x_S$, it is referred to as a prior because it serves as the prior distribution for the subsequent analysis of the target data.} 

{\color{black}Most Bayesian extrapolation methods assess similarity between source and target populations using outcome data and adjust the degree of borrowing according to the observed commensurability between populations. However, differences between populations may also be partially explained by measured baseline covariates, including prognostic factors and treatment-effect modifiers. From a causal-inference perspective, such covariates may influence the plausibility of exchangeability between the source and target populations. This has motivated the development of {\it Bayesian causal inference} approaches, which incorporate covariate information to adjust for observed population differences before borrowing information from the source population. These methods are discussed within the relevant borrowing frameworks below.}

\subsubsection{Adaptive down-weighting of data from the source population}

Approaches in this category incorporate data from source populations into current target population studies by adaptively weighting the $x_S$ according to its commensurability with $x_T$.

\paragraph{Power priors}

One approach that has received substantial attention is the power prior\cite{RN430,RN316,RN162,RN167,RN397,RN333,RN418,RN310,RN790,RN665,RN64,Baumann03072025,RN409,RN146,RN23,RN170,RN784,RN104,RN196,RN338,RN470,RN918,RN328,RN311,RN325,RN82,RN61,RN91,RN359,RN357,RN309,RN330,RN340,RN218,wang2019propensity,baron2022bayesian,RN644,RN6,RN83}. 
Power priors construct the prior distribution for the target population parameter by updating an initial prior with discounted information from the source population. Specifically, the likelihood of $x_S$ is down-weighted by raising it to a power $a_0 \in [0, 1]$. Assuming parameters are consistent across populations, such that $\theta_T = \theta_S = \theta$, let $L(\theta \mid x_S)$ denote the likelihood of $x_S$, {\color{black}$\pi_0(\theta)$ represent the initial prior for $\theta$, and $\pi(\theta,a_0|x_S)$ denote the joint power prior distribution for $(\theta,a_0)$ given $x_S$. The power prior is defined as 
\begin{equation}
	\pi(\theta,a_0|x_S)=C(a_0)L(\theta|x_S)^{a_0}\pi_0(\theta)\pi(a_0),  
	\label{modified pp}
\end{equation}
where $C(a_0)=1/{\int L(\theta|x_S)^{a_0}\pi_0(\theta)\text{d}\theta}$ is the normalising constant that depends on $a_0$\cite{RN328}, and $\pi(a_0)$ is the prior distribution for $a_0$, reflecting uncertainty about the commensurability of parameters between the source and current data.} Ibrahim and Chen (2000)\cite{RN430} suggest a Beta, truncated Gamma, or Normal prior for $a_0$. 
They further extend the power prior \eqref{power prior} to accommodate multiple source datasets and to generalised linear, proportional hazards, and cure rate models. 

Data $x_T$ from the new trial are used to update \eqref{power prior} via Bayes' theorem to obtain the posterior distribution of $(\theta, a_0)$ given $(x_S, x_T)$. Both datasets contribute to learning about $a_0$, determining the effective contribution of $x_S$ to the marginal posterior of $\theta$. The underlying motivation is that, when $x_S$ and $x_T$ are commensurate, the posterior distribution of $a_0$ should favour values close to 1, leading to substantial borrowing. Conversely, when the datasets are discordant, the posterior distribution should assign more mass to values of $a_0$ near 0, thereby discounting the source data. In practice, the extent to which this adaptive behaviour occurs depends on the amount and type of data available and the prior distributions. Consequently, the choice and modelling of $a_0$ is critical for constructing a reasonable power prior. {\color{black}Because the degree of borrowing is determined using the observed data, adaptive power priors share some similarities with empirical Bayes approaches. Therefore, atypical target datasets may occasionally influence the borrowing weight and favour borrowing from source datasets with similar characteristics. Therefore, the choice and modelling of $a_0$, together with evaluation of the operating characteristics of the procedure, are critical for constructing a reasonable power prior.}

{\color{black}Historically, the power prior was first proposed in the form of the hierarchical power prior \cite{RN430}, denoted $\pi^{\text{PP}}(\theta,a_0|x_S)$, in which
\begin{equation}
	\pi^{\text{PP}}(\theta,a_0|x_S)\propto L(\theta|x_S)^{a_0}\pi_0(\theta)\pi(a_0)\, ,
	\label{power prior}
\end{equation}
where the normalising constant $C(a_0)$ is omitted.} As discussed by Neuenschwander et al. (2009)\cite{RN328}, this formulation violates the likelihood principle when $a_0$ is treated as an unknown parameter because the omitted normalising constant depends on $a_0$. Consequently, the formulation is valid when $a_0$ is fixed, but not when a prior distribution is assigned to $a_0$. For this reason, many subsequent approaches recommend selecting a single value of $a_0$ based on the degree of similarity between $x_S$ and $x_T$, rather than assigning a prior distribution $\pi(a_0)$.   
Ibrahim et al. (2015)\cite{RN316} propose four selection criteria and introduce a partial discounting power prior, which allows borrowing of external information only through parameters shared between the target and source populations. Gravestock and Held (2017)\cite{RN162} adopt an empirical Bayes approach, selecting $a_0$ by maximising its marginal likelihood, and later\cite{RN167} extend this framework to multiple source datasets. Fei{\ss}t et al. (2020)\cite{RN397} impose an upper bound on $a_0$ to ensure type I error control. Alt et al. (2023)\cite{RN333} introduce the scale-transformed power prior and its variations for borrowing across outcome types, selecting $a_0$ to optimise model fit using the deviance information criterion. 

Pan et al. (2017)\cite{RN418} propose the calibrated power prior, determining $a_0$ using the Kolmogorov-Smirnov statistic. Nikolakopoulos et al. (2018)\cite{RN310} and Psioda and Xue (2020)\cite{RN790} use Box's Bayesian $p$-value\cite{box1980sampling} to quantify prior-data conflict and adjust $a_0$. 
Belay et al. (2021)\cite{RN665} extend the calibrated power prior framework to time-to-event endpoints, while Kwiatkowski et al. (2024)\cite{RN64} generalise it by introducing subject- and interval-specific $a_0$. 
Baumann et al. (2025)\cite{Baumann03072025} extend the calibrated power prior to basket trials. 

Even with such a fixed $a_0$, the degree of borrowing is not inferred within a probabilistic model for population differences. A related limitation is the potential lack of robustness, as the effective contribution of the source data scales with $a_0 n_S$. Hence, for large $n_S$, the source data may dominate the posterior even under conflict. 

Other approaches define $a_0$ as a function of similarity between populations. Liu (2018)\cite{RN409} uses a monotonic function of the $p$-value from an equivalence test comparing $\theta_T$ and $\theta_S$, while Lin and Slate (2026)\cite{RN146} base $a_0$ on the $z$-score for the difference between the sample means of $x_S$ and $x_T$. 
Shi et al. (2023)\cite{RN23} use the overlap area between the distributions of target and source populations, whereas Qian et al. (2025)\cite{RN170} use similarity in propensity score distributions. 
Basu et al. (2020)\cite{RN784} calibrate $a_0$ using a PK/PD model to quantify similarity. 
Bennett et al. (2021)\cite{RN104} propose probability-weighted and equivalence probability-weighted methods that quantify agreement between populations within predefined equivalence bounds, and Peng et al. (2023)\cite{RN196} extend this approach to multiple source datasets. 
Additionally, three papers\cite{RN338, RN470, RN918} decompose $a_0$ into maximum allowable borrowing and commensurability components.

{\color{black}To address the violation of the likelihood principle in \eqref{power prior}, the {\it modified power prior}\cite{RN328,RN311} incorporates the normalising constant $C(a_0)$ and is defined by Equation \eqref{modified pp}.} 
Building on this, Chen et al. (2011)\cite{RN325} extend the modified power prior to accommodate multiple sources datasets as well as binary and normally distributed outcomes. Banbeta et al. (2019, 2022, 2024)\cite{RN82,RN61,RN91} extend the modified power prior to variants for multiple source datasets, linear regression models with covariate adjustment, and count data. 
Ye et al. (2022)\cite{RN359} establish the theoretical optimality of the modified power prior, showing that it minimises a weighted Kullback-Leibler divergence, and generalise it to multiple source datasets. Wang et al. (2025)\cite{RN357} propose the modified borrowing-by-parts power prior, which separately down-weights mean and variance components, and introduce a robust version that fully discounts external variance.  
Hobbs et al. (2011)\cite{RN311} modify the modified power prior by using a normal prior for $\theta$ with mean $\theta_S$ and precision, termed the location-commensurate power prior. 
Tran (2023)\cite{RN309} extends this by defining boundary conditions for $a_0$ based on whether a desired power level is achieved for specified type I error rate.

For complex models, a key limitation of the modified power prior is the computational burden of repeatedly evaluating $C(a_0)$ during Markov chain Monte Carlo sampling. Han et al. (2023)\cite{RN330} introduce an efficient importance sampling approach that substantially reduces computational cost.  
Han et al. (2025)\cite{RN340} further address this for multiple source datasets. 

While adaptive behaviour is often highlighted as a strength of power prior approaches, specification of the prior distribution for $a_0$ remains a challenging and potentially influential decision. Although standard choices such as Beta priors are frequently adopted, it is not always clear how to justify a particular form or assess its informativeness in practice \cite{rover2025meta}.

{\color{black}Several {\it Bayesian causal inference} methods have been proposed for constructing the power prior. These methods incorporate baseline covariate information when determining the degree of borrowing from the source population and differ in how source-target population differences are quantified and incorporated into the power parameter. A first group of approaches uses propensity score methods.} Lin et al. (2019)\cite{RN218} combine propensity score matching with the power prior, where each matched observation is assigned a power parameter determined by its propensity score. Wang et al. (2019)\cite{wang2019propensity} extend this approach to multiple source datasets by stratifying observations and assigning stratum-specific power parameters according to the overlap between propensity score distributions. Baron et al. (2022)\cite{baron2022bayesian} further refine this framework by defining the power parameter as a function of stratum-specific mean differences, while Baron et al. (2025)\cite{RN644} generalise the approach to separately down-weight mean and variance components. {\color{black}A second group of methods relies on weighting approaches.} Wang et al. (2022)\cite{RN6} use inverse probability of treatment weighting to reweight source population data and subsequently extend this framework by incorporating a Bayesian bootstrap to account for uncertainty arising from both weighting and sampling\cite{RN83}.

{\color{black}Although these methods adjust for observed differences in baseline covariate distributions between populations, borrowing based on absolute clinical outcomes may remain susceptible to biases arising from study-level differences, including regression-to-the-mean effects. In particular, studies enrolling patients with different baseline disease severity may exhibit different magnitudes of spontaneous improvement that may not be fully captured through covariate adjustment alone. Consequently, when suitable control data are available, borrowing comparative treatment effects derived from treatment-control comparisons may be more robust than borrowing absolute response outcomes. This consideration is particularly relevant when determining whether borrowing should be based on data from a single treatment or control arm, or on information from both treatment and control arms of the source study. 
}

{\color{black}Power priors provide a flexible and intuitive framework for borrowing information from source data through a discounting parameter that directly controls the contribution of external evidence. Their principal advantage is simplicity and interpretability. However, their performance depends critically on the specification of the borrowing parameter. Fixed power priors may be sensitive to source-target heterogeneity and can lead to over-borrowing when population differences exist. Adaptive power priors partially address this limitation by reducing borrowing when prior-data conflict is detected, although they may themselves become sensitive to atypical target datasets. Hence, power priors are particularly attractive when source population information is believed to be relevant but uncertainty remains regarding the appropriate degree of borrowing.}

\paragraph{Commensurate prior}

A similar Bayesian model for $x_S$ and $x_T$ underlies the derivation of the commensurate prior for $\theta_T$ \cite{RN311,RN6,RN158,RN52,RN742,RN203,RN459,RN455,RN363,RN754,RN181}. 
The commensurate prior approach modifies the likelihood to account for differences between $\theta_T$ and $\theta_S$.  
In this framework, suppose that $\theta_T \neq \theta_S$, while $\mathbb{E}(\theta_T)=\theta_S$ and $\text{Var}(\theta_T|\theta_S) = \tau^{-1}$. The commensurate prior for $\theta_T$ given $x_S$ and $\theta_S$ is defined as 
\begin{equation}
	\pi^{\text{CP}}(\theta_T,\tau|x_S)\propto \Big[\int L(\theta_S|x_S)\pi(\theta_T|\theta_S,\tau)\pi_0(\theta_S)\text{d}\theta_S\Big]\pi_0(\theta_T)\pi(\tau),
	\label{commensurate prior}
\end{equation}
where $\pi(\tau)$ represents the degree of borrowing between source and current data. 
This formulation is closely related to the bias-allowance structure described by Welton et al. (2012, Section 11.3)\cite{welton2012evidence}.
Given $x_T$, the posterior density for $(\theta_T, \tau)$ is proportional to \eqref{commensurate prior} multiplied by $L(\theta_T|x_T)$. If $x_S$ and $x_T$ are inconsistent (i.e., $\tau \rightarrow 0$), $x_S$ are effectively discarded, and the marginal posterior for $\theta_T$ approaches that obtained by updating the initial prior $\pi_0(\theta_T)$ using $x_T$ alone. Conversely, when the datasets are highly consistent ($\tau \rightarrow \infty$), the marginal posterior for $\theta_T$ converges to that obtained by pooling $x_T$ and $x_S$ under the assumption $\theta_T = \theta_S$. Hence, the choice of the parameter $\tau$ is critical.

Hobbs et al. (2012, 2013)\cite{RN158,RN52} recommend defining $\pi(\tau)$ in \eqref{commensurate prior} either as a conditionally conjugate prior or as a spike-and-slab prior. Alternatively, an empirical Bayes approach replaces $\tau$ with its marginal maximum likelihood estimate\cite{RN158}. Hobbs et al. (2013)\cite{RN52} apply the commensurate prior framework to incorporate control data from a source population in an adaptive randomised controlled trial, where the randomisation ratio is updated sequentially based on the information contributed by $x_S$. 
The commensurate prior approach is further generalised to accommodate a variety of data types. 
Hobbs et al. (2012)\cite{RN158} illustrate commensurate priors with binary, survival, and count data. Hobbs et al. (2011)\cite{RN311} adapt the commensurate prior for normally distributed outcomes, introducing the location commensurate prior, which shrinks the current treatment effect $\theta_T$ toward the corresponding estimate from $x_S$. They further extend this framework to the location-scale commensurate prior, in which borrowing depends jointly on the similarity of means and variances.

Murray et al. (2015)\cite{RN742} introduce the generalised mixture commensurate prior, enabling feature-specific borrowing through a hierarchical spike-and-slab structure. 
Gajewski et al. (2016)\cite{RN203} extend the commensurate prior to multiple source datasets. 
Zheng and Wason (2022)\cite{RN459} and Zheng et al. (2023a, 2023b)\cite{RN455,RN363} adapt this prior for basket trials by introducing mixture priors on $\tau$, allowing borrowing to vary according to the commensurability between baskets. Ouma et al. (2022)\cite{RN754} generalise this approach by enabling differential borrowing from treatment and control arms based on group-wise mean similarities. Khanal et al. (2025)\cite{RN181} propose a clustering-based commensurate prior method for basket trials, which groups baskets according to the commensurability then facilitates adaptive borrowing within each cluster. Wang et al. (2022)\cite{RN6} propose a {\it Bayesian causal inference} method that combines the commensurate prior with propensity score methods to balance covariates before adaptively borrowing.

{\color{black}Commensurate priors differ from power priors in that they directly model the similarity between source and target parameters rather than borrowing by discounting the contribution of source data through a borrowing parameter. This often provides an intuitive interpretation of borrowing and may offer additional protection against prior-data conflict. However, the notion of commensurability remains difficult to quantify in practice and often requires careful clinical justification.}

\paragraph{Bayesian hierarchical model and meta-analytic predictive prior}

The Bayesian hierarchical model provides a Bayesian framework for combining information across multiple source populations to construct an informative prior for $\theta_T$\cite{RN325,RN91,RN549,RN139,RN94,RN217,RN18,RN807,RN432,RN159,RN150,RN599,RN33,RN600,RN554,RN225,RN448,RN40,RN49,RN661,RN995,RN344,RN103}. 
The meta-analytic predictive prior\cite{RN139,RN91,RN325,RN159,RN150,RN49,RN661,RN33,RN344,RN103} arises naturally from the Bayesian hierarchical model and provides a coherent framework for synthesising evidence across studies by modelling study-specific effects as exchangeable draws from a common distribution governed by hyperparameters representing the overall mean and between-study heterogeneity. 
Accordingly, we consider the meta-analytic predictive prior and the Bayesian hierarchical model under the same methodological category.

The Bayesian hierarchical model for normally distributed outcomes is given by
\begin{align}
	&X_{Sh}|\theta_{Sh}\sim N(\theta_{Sh}, \sigma_{Sh}^2),\quad h=1,\dots,H,\nonumber\\
	&\theta_{S1},\dots,\theta_{SH},\theta_T|\theta,\nu\sim N(\theta,\nu^2),\nonumber\\
	&\theta\sim \pi(\theta),\quad \nu\sim \pi(\nu),
	\label{Bayesian meta-analysis model}
\end{align}
where $\sigma_{S1},\dots,\sigma_{SH}$ are assumed known. When the between-trial variance $\nu^2$ is known, the posterior distribution of $\theta$ given the source population data is 
$$\theta|x_{S1},\dots,x_{SH},\nu\sim N\left(\frac{\sum w_h x_{Sh}}{\sum w_h},\, \frac{1}{\sum w_h}\right),$$
where $w_h = 1/(\sigma_{Sh}^2 + \nu^2)$. 

The prior distribution of $\theta_T$ is given by its posterior predictive distribution conditional on the source data $$\theta_T|x_{S1},\dots,x_{SH},\nu\sim N\left(\frac{\sum w_h x_{Sh}}{\sum w_h},\, \frac{1}{\sum w_h}+\nu^2\right).$$ 
When $H=1$, the Bayesian hierarchical model is identical to the power prior in \eqref{power prior} if $a_0=(2\nu^2/\sigma_{S1}^2+1)^{-1}$ \cite{chen2006relationship,rover2025meta}, and to the commensurate prior in \eqref{commensurate prior} if $\tau^{-1}=2\nu^2$ \cite{RN158}.

Thall et al. (2003)\cite{RN549} introduce the Bayesian hierarchical model for within-trial borrowing across disease subtypes in phase II oncology studies, enabling partial pooling while maintaining subtype-specific decision rules. Neuenschwander et al. (2010)\cite{RN139} recommend sensitivity assessment for the prior on $\nu$ in fully Bayesian analyses.  
Gsteiger et al. (2013)\cite{RN94} and Banbeta et al. (2024)\cite{RN91} derive meta-analytic predictive priors for count data, while Chen et al. (2011)\cite{RN325} propose a unified meta-analytic predictive framework for normal and binary outcomes. Cuffe (2011)\cite{RN217} introduces an adaptive borrowing scheme based on the estimated between-trial variance, which discounts information from the source population by restricting the effective sample size of the external dataset. 

Several studies extend the Bayesian hierarchical model and meta-analytic predictive to specific data structures. Han et al. (2017)\cite{RN18} propose a covariate-adjusted Bayesian hierarchical model for time-to-event outcomes. 
Chu and Yuan (2018)\cite{RN807} develop a calibrated Bayesian hierarchical model for basket trials, in which the between-study variance is calibrated via simulation under scenarios of complete homogeneity and strong heterogeneity. Chi et al. (2024)\cite{RN432} generalise this approach to multiple endpoints. 
Bertsche et al. (2019)\cite{RN159} and Roychoudhury and Neuenschwander (2020)\cite{RN150} apply the meta-analytic predictive prior to survival models. 
Zhao and Koschmann (2020)\cite{RN599} introduce a joint Bayesian hierarchical model for mixed endpoints. 
Li et al. (2023)\cite{RN33} extend the Bayesian hierarchical model to study-level longitudinal data, applying the meta-analytic predictive prior separately at each time point. 

Other extensions focus on modelling correlation and heterogeneity. R\"{o}ver and Friede (2020)\cite{RN600} propose a Bayesian shrinkage estimator within the normal-normal hierarchical framework, which shrinks the parameter for the target population toward the overall parameter estimated from both target and source populations. Jin et al. (2020)\cite{RN554} introduce a correlated Bayesian hierarchical model that incorporates pairwise similarities between indications using divergence measures between their posterior distributions. Holzhauer (2020)\cite{RN225} employs shrinkage priors with shift parameters to account for bias between source and current data, whereas Fujikawa et al. (2020)\cite{RN448} assign borrowing weights based on posterior similarity quantified via the Jensen-Shannon divergence. Collignon et al. (2020)\cite{RN40} advocate modelling multiple variance components to capture within-patient, between-centre, and across-study variability. 

Further innovations include a semiparametric meta-analytic predictive prior using a Dirichlet process for $\nu$ (Hupf et al., 2021)\cite{RN49}, an elastic meta-analytic predictive prior that discounts borrowing according to data congruence (Zhang et al., 2021)\cite{RN661}, and an optimal Bayesian hierarchical model selecting the prior for $\nu$ by maximising a utility function balancing power and type I error (Jiang et al., 2021)\cite{RN995}.

Two papers consider Bayesian causal inference methods within the Bayesian hierarchical model framework. 
Liu et al. (2021)\cite{RN344} use propensity scores to trim and stratify $x_S$ and $x_T$ into homogeneous subgroups based on baseline covariates, constructing stratum-specific meta-analytic predictive priors that account for residual heterogeneity. Hupf et al. (2024)\cite{RN103} extend this approach by assigning independent meta-analytic predictive priors to individual covariate effects, allowing the degree of borrowing to depend on the consistency across studies rather than on overall outcome similarity.

Note that borrowing under the Bayesian hierarchical model framework can be interpreted through study weights analogous to inverse-variance weights in frequentist meta-analysis. In the Bayesian setting, these weights arise naturally for both overall and shrinkage estimates, including the meta-analytic predictive prior, and quantify the contribution of each study to the resulting estimate. This perspective helps address the common concern that source data may overwhelm the target, particularly when the source study is large. In fact, the posterior mean shrinkage weights ensure that the contribution of the current data is bounded below, implying that 
borrowing from external data is inherently limited under the hierarchical model framework \cite{RN444}. 

{\color{black}Meta-analytic predictive priors and Bayesian hierarchical models provide a natural framework for borrowing information across multiple studies or subpopulations by modelling exchangeability through a common hierarchical structure. These methods are particularly attractive when several source studies are available and between-study heterogeneity can be estimated, as well as in basket trials where information is shared across subgroups. Their principal advantage is the ability to adapt borrowing according to observed heterogeneity. In settings involving multiple source datasets, information is borrowed differentially, with studies that are more similar to the target population contributing more to the final inference than studies exhibiting greater heterogeneity. However, the performance of these methods depends on the validity of the exchangeability assumption and the specification of heterogeneity parameters. To address these limitations, robust meta-analytic predictive priors have been developed to reduce borrowing in the presence of prior-data conflict, typically by combining an informative exchangeable component with a weakly informative or non-exchangeable component in a mixture prior.}

\paragraph{Mixture prior}

Mixture prior methods\cite{RN995,RN909,RN326,RN318,RN123,RN73,RN673,RN361,RN922,RN887,RN343,RN1162,RN8,RN9,RN933,RN16,RN508,RN882,RN481,RN535,RN472,RN594,RN908,RN552,RN81,RN912,RN883} specify the prior for $\theta_T$ as 
\begin{equation}
	\pi(\theta_j|x_S)=w_j\pi_1(\theta_j)+(1-w_j)\pi_2(\theta_j),\ \text{with}\ j= S1, \dots, SH, T,
	\label{mixture prior}
\end{equation}
where $\pi_1(\cdot)$ represents an informative prior derived from $x_S$, and $\pi_2(\cdot)$ denotes a non-informative (or weakly informative) prior diluting the influence of $x_S$, thereby producing a prior with heavier tails. 

Hsiao et al. (2007)\cite{RN909} propose that the mixture weight $w_T$ for the target population be pre-specified by the regulatory authority in the new region for bridging studies between regions. 
Hobbs et al. (2011)\cite{RN311} generalise this approach by using a mixture of multiple location-scale commensurate priors. 
Yang et al. (2023)\cite{RN326} develop the self-adjusting mixture prior, in which mixture weights $w_j$ are adaptively determined based on the degree of prior-data conflict, while Calderazzo et al. (2024)\cite{RN318} link the mixture weights to type I error control through a linear combination of frequentist and Bayesian decision thresholds.

Many methods combine the Bayesian hierarchical model with a mixture prior structure. Schmidli et al. (2014)\cite{RN123} introduce the robust meta-analytic predictive prior, defined as a mixture of the meta-analytic predictive prior and a vague conjugate prior. 
Zhao and Ma (2024)\cite{RN73} and Zhang et al. (2023)\cite{RN673} propose empirical Bayes variants of the robust meta-analytic predictive prior, in which mixture weights are determined using tail probabilities or prior predictive $p$-values, respectively. Di Stefano et al. (2023)\cite{RN361} adapt the robust meta-analytic predictive prior framework to pharmacokinetic settings, incorporating physiologically based priors for treatment and control arms. 
Neuenschwander et al. (2016)\cite{RN922} develop the exchangeable-nonexchangeable model, which allows partial pooling by modelling subgroup-specific parameters as a mixture of exchangeable and non-exchangeable components. 
Chen and Hsiao (2023)\cite{RN887} extend the exchangeable-nonexchangeable model by dynamically updating exchangeability weights during interim analyses.   
Li et al. (2016)\cite{RN343} propose the empirical meta-analytic predictive prior, which estimates the between-study variance empirically from the current data $x_T$ rather than specifying fixed mixture weights. Liu et al. (2017)\cite{RN1162} extend the Bayesian hierarchical model through a hierarchical mixture model with two exchangeable components representing different degrees of variability.
Ohigashi et al. (2022)\cite{RN8} incorporate horseshoe priors within the Bayesian hierarchical model to achieve adaptive shrinkage,  
and Ohigashi et al. (2025)\cite{RN9} further generalise this approach to include three alternative shrinkage priors.

Four works\cite{RN933,RN16,RN508,RN882} consider Bayesian causal inference methods for constructing mixture priors. Zhu et al. (2024)\cite{RN933} extend the robust meta-analytic predictive prior by integrating propensity score-based, stratum-specific priors with a vague component. Callegaro et al. (2024)\cite{RN16} adjust the source population component of the mixture prior using three causal inference approaches: inverse probability of treatment weighting, G-computation, and a doubly robust estimator. Building on Yang et al. (2023)\cite{RN326}, Zhao et al. (2025)\cite{RN508} refine the self-adjusting mixture prior by incorporating propensity score matching and inverse probability of treatment weighting to align covariate distributions between source and current data. Psioda et al. (2025)\cite{RN882} propose an inverse probability-weighted mixture prior, combining propensity score-based weighting with the power prior within the mixture-prior framework.

Ten methods\cite{RN995,RN481,RN535,RN472,RN594,RN908,RN552,RN81,RN912,RN883} construct the mixture priors using non-parametric Bayesian modelling. These methods proceed in two stages: (1) classifying datasets into subgroups based on the similarity between external and current datasets, and (2) adaptively borrowing information from identified subgroups via hierarchical or covariance structures. 

{\it Bayesian clustering} is the most commonly used non-parametric approach. Leon-Novelo et al. (2012)\cite{RN481} introduce a non-exchangeable product partition model that enables data-driven grouping through a modified cohesion function incorporating covariate-based similarity. Guo et al. (2016)\cite{RN535} develop a tree-based Bayesian hierarchical model framework for multi-regional clinical trials, where borrowing strength is determined by a hierarchical clustering of regions.  
Chen and Lee (2019)\cite{RN472} propose a Bayesian classification and information-sharing model that classifies subgroups into high- and low-response clusters using latent variables, and later extend this work by dynamically determining the number of clusters through a Dirichlet process prior\cite{RN594}. Li and Yuan (2025)\cite{RN908} adopt a similar strategy for hierarchical borrowing across adult and paediatric populations. 
Jiang et al. (2021)\cite{RN995} apply Bayesian clustering to group tumour types into sensitive and insensitive clusters before implementing an optimal Bayesian hierarchical model within each cluster. Daniells et al. (2023)\cite{RN552} modify the exchangeable-nonexchangeable model by introducing a two-step clustering-based weighting scheme that excludes heterogeneous baskets and computes similarity-based borrowing weights using Hellinger distances.

Bayesian causal inference methods are also considered within Bayesian clustering. Zhou and Ji (2021)\cite{RN81} and Pan et al. (2022)\cite{RN912} employ Bayesian Additive Regression Trees. 
Long et al. (2024)\cite{RN883} propose a Bayesian hierarchical clustering tree that uses propensity score-based similarity to group datasets. 

{\color{black}Mixture priors provide a flexible framework for robust information borrowing by combining multiple prior components that represent different assumptions regarding source-target similarity. Their principal advantage is the ability to reduce the risk of over-borrowing while still retaining efficiency gains when the source and target populations are genuinely similar. Consequently, mixture priors are particularly attractive when there is uncertainty regarding exchangeability or when regulatory concerns exist about excessive reliance on external information, making them popular in applications such as basket trials and multi-region clinical trials. However, their performance depends on the specification of the mixture components and their associated weights. In practice, determining appropriate mixture weights may require expert judgement, empirical evidence, or data-adaptive procedures, and different choices can lead to different borrowing behaviour.}

\paragraph{Bayesian model averaging}

Bayesian model averaging\cite{RN995,RN362,RN374,hobbs2018bayesian,RN910,RN539,bean2023bayesian,RN498,RN437,RN207,RN484,RN315,RN141,RN160,RN562,RN551,RN442,RN1097} is a general framework for constructing informative priors for $\theta_T$ by accounting for model uncertainty through averaging over multiple candidate models weighted by their posterior probabilities. 
Kaizer et al. (2018a, 2018b)\cite{RN362,RN374} introduce multisource exchangeability models to systematically evaluate potential exchangeability structures among source and current datasets. 
When multiple source datasets are available, multisource exchangeability models construct a weighted average over all possible exchangeability configurations. The mixture prior in \eqref{mixture prior} can be interpreted as a form of Bayesian model averaging, in which inference is averaged between an informative borrowing model and a non-informative model excluding external data.

Hobbs and Landin (2018)\cite{hobbs2018bayesian} extend the multisource exchangeability model to basket trials by evaluating all possible pairwise exchangeability configurations among baskets, and propose empirical Bayes implementations to improve computational efficiency and control shrinkage. Asano and Hirakawa (2020)\cite{RN910} simplify Bayesian model averaging based designs by focusing on two interpretable models, one assuming complete homogeneity and another allowing heterogeneity, estimating model weights via empirical Bayes. Liu et al. (2022)\cite{RN539} develop the local multisource exchangeability model for phase II basket trials. 
Bean et al. (2023a, 2023b)\cite{bean2023bayesian,RN498} apply Bayesian model averaging to multi-regional clinical trials, first addressing continuous outcomes and later extending the approach to time-to-event data. 
Bean et al. (2024)\cite{RN437} further expand this framework by jointly modelling time-to-event and longitudinal outcomes. Ji and Wolfson (2023)\cite{RN207} propose a data-driven multisource exchangeability model that integrates adaptive source selection and clustering to reduce the computational complexity of enumerating exchangeability configurations. 
Wolf et al. (2024)\cite{RN484} extend Bayesian model averaging to multiple outcomes, allowing secondary endpoints to inform exchangeability. Alt et al. (2024)\cite{RN315} introduce the latent exchangeability prior, linking subject-level exchangeability to Bayesian model averaging through model averaging over uncertainty in exchangeability configurations. Wei et al. (2024)\cite{RN141} establish a conceptual link between Bayesian model averaging and the power prior, interpreting subgroup-specific power parameters as posterior probabilities of exchangeability. Jiang et al. (2021)\cite{RN995} apply Bayesian model averaging within the Bayesian hierarchical model in an oncology setting, treating each configuration of tumour-type sensitivity as a separate model.

Five papers\cite{RN160,RN562,RN551,RN442,RN1097} construct Bayesian causal inference approaches within the Bayesian model averaging frameworks. Kotalik et al. (2021)\cite{RN160} extend the multisource exchangeability model using the potential outcomes framework, allowing borrowing only when differences in marginal treatment effects are attributable to covariate imbalance rather than true heterogeneity. Boatman et al. (2021)\cite{RN562} apply multisource exchangeability models to regression-based causal effect estimators to infer population average treatment effects. Wei et al. (2024)\cite{RN551} propose a propensity score-adjusted multisource exchangeability model that modifies the likelihood of each source dataset to account for baseline differences. Haine et al. (2024)\cite{RN442} introduce a semi-supervised mixture multisource exchangeability model that uses propensity score-based clustering to identify source datasets most comparable to the target population, while Haine et al. (2024)\cite{RN1097} extend this framework to interim futility monitoring. 

{\color{black}Similar to mixture priors, Bayesian model averaging is particularly attractive when there is uncertainty regarding exchangeability. Its principal advantage is that it formally incorporates model uncertainty into the analysis rather than conditioning inference on a single borrowing model. Therefore, Bayesian model averaging may provide greater robustness when there is uncertainty regarding the validity of extrapolation assumptions. However, its performance depends on the choice of candidate models and prior model probabilities, and the resulting inference may be sensitive when the available data provide limited information for distinguishing between competing models. In practice, Bayesian model averaging is particularly attractive when multiple plausible borrowing assumptions exist and there is no strong justification for selecting a single model a priori.}

\paragraph{Other Bayesian methods}

In addition, five methods\cite{RN352,RN317,RN306,RN237,RN914} construct informative priors for $\theta_T$ directly from source datasets. Jin and Yin (2021)\cite{RN352} propose the unit information prior, which adaptively borrows from multiple sources by defining the prior through Fisher's information. 
Jiang et al. (2023)\cite{RN317} develop the elastic prior, which adjusts the variance of the source-data posterior according to a congruence measure, allowing borrowing to vary smoothly from complete to negligible as similarity decreases. Warren et al. (2024)\cite{RN306} introduce the scaled Gaussian kernel density estimation prior, which approximates the source-data posterior using kernel density estimation to serve as the base prior for $\theta_T$. Brizzi et al. (2023)\cite{RN237} propose a model-informed prior, generating the prior-predictive distribution for the control effect via simulations based on source data. 
Wang and Tiwari (2025)\cite{RN914} present the Bayesian bootstrap with covariate balancing dynamic borrowing approach.

\subsubsection{Non-adaptive down-weighting of data from the source population}

Twenty methods\cite{RN311,RN430,RN369,RN715,RN1419,RN499,RN138,RN428,RN816,RN651,RN2011,RN168} use existing data from a source population to formulate an informative prior for $\theta_T$, down-weighting these data in a non-adaptive, pre-specified manner. 

Ten methods\cite{RN430,RN311,RN369,RN715,RN1419,RN499,RN138,RN428} propose power priors with fixed $a_0$. In contrast to adaptive down-weighting power priors, the value of $a_0$ is not determined from the observed $x_S$ and $x_T$. Several studies\cite{RN430,RN369,RN715} employ such power priors without specifying a formal method for determining $a_0$. 
Others provide approaches for choosing $a_0$. Gandhi et al. (2012)\cite{RN1419} recommend sensitivity analyses to assess $a_0$'s impact.   Ibrahim et al. (2012)\cite{RN499} propose a partial borrowing power prior that limits borrowing to parameters shared by the source and current models. Yuan et al. (2022a, 2022b)\cite{RN138,RN428} extend this concept through borrowing-by-parts power priors. 
They also propose conditional borrowing power priors, where borrowing occurs only if current observations fall within a predefined subset, and develop a robust version using a mixture prior framework.

Additionally, Hobbs et al. (2011)\cite{RN311} propose a variation of the commensurate prior where $\tau$ in \eqref{commensurate prior} is fixed. Within the Bayesian hierarchical model framework, Schoenfeld et al. (2009)\cite{RN816} implement non-adaptive down-weighting by assigning pre-specified values to $\theta$ and $\nu$ in \eqref{Bayesian meta-analysis model}. Six methods\cite{RN651,RN2011,RN168} construct informative priors for $\theta_T$ by incorporating source data while applying non-adaptive, pre-specified down-weighting to the variance.

{\color{black}Non-adaptive Bayesian borrowing methods rely on prespecified assumptions regarding source-target similarity and employ a fixed borrowing mechanism. They are simple to implement and can provide substantial efficiency gains when these assumptions are valid. However, they offer limited protection against prior-data conflict and source-target heterogeneity, and are therefore most appropriate when strong external evidence supports the borrowing assumptions.}

\subsection{Frequentist methods}

{\color{black}Frequentist borrowing methods pursue objectives similar to those of Bayesian approaches but achieve them through different mechanisms. Rather than incorporating information from source populations through prior distributions, these methods typically rely on weighting, shrinkage, penalisation, or the combination of test statistics.}

A total of 25 frequentist methods\cite{RN111,RN519, quan2014multi, RN206, RN396, RN80, RN265, RN789, RN697, RN492, RN147, RN567, RN605, RN1544, RN1473, RN920, RN786, RN900, RN56, RN672, RN63, RN885, RN1805} were identified and grouped into three broad categories: joint modelling of source and target data, combination of data via weighted test statistics, and assessment of parameter consistency between populations before borrowing.

\subsubsection{Joint model incorporating data across populations}

Thirteen methods\cite{RN111,RN519, quan2014multi, RN206, RN396, RN80, RN265, RN789, RN697, RN492, RN147, RN567, RN605} synthesise data from source and target populations using a frequentist random-effects model.

The first type defines the estimator of the control-arm effect as a weighted average of the effect estimates from the current and source control arms \cite{RN111, RN519, quan2014multi,RN206, RN396}. Let $\hat{\theta}_{T0}$ and $\hat{\theta}_{S0}$ denote the effect estimates for the current and source control arms, respectively. The estimator of the control-arm effect is given by 
\begin{equation*}
	\hat{\theta}_0=\lambda \hat{\theta}_{T0}+(1-\lambda)\hat{\theta}_{S0}
\end{equation*}
where $\lambda$ is a weight parameter that balances the contributions from the current and historical controls. 
These methods consider different strategies for selecting $\lambda$. 
Thall and Simon (1990)\cite{RN111} use a univariate random-effects meta-analysis, in which $\lambda$ is chosen to minimise the variance of the target treatment effect $\hat{\theta}=\hat{\theta}_{T1}-\hat{\theta}_0$, where $\hat{\theta}_{T1}$ denotes the effect estimate for the current experimental arm. 
Quan et al. (2013)\cite{RN519} determine $\lambda$ based on within-region variance and sample size in multi-regional clinical trials.  
Quan et al. (2014)\cite{quan2014multi} extend this framework using a James-Stein shrinkage estimator. Galwey (2017)\cite{RN206} further refines this by introducing a bias-variance trade-off model, while Chu and Yi (2021)\cite{RN396} select $\lambda$ to minimise the mean squared error.

Other methods propose two-stage approaches that adjust source and current data before determining a shrinkage estimator \cite{RN80, RN265, RN789, RN697}. Li and Jemielita (2023)\cite{RN80} adjust data using propensity score weighting or matching combined with partial bias correction. Tan et al. (2022)\cite{RN265} address residual bias between populations using regression-based methods.  Gamalo et al. (2023)\cite{RN789} down-weight source data using a bounded function of the dissimilarity. 
Holzhauer and Adewuyi (2023)\cite{RN697} introduce a model-based prognostic borrowing method that incorporates predictions from source data as covariates in the regression model of the new trial before estimating treatment effects.

Four papers\cite{RN492, RN147, RN567, RN605} consider a penalised likelihood framework to dynamically integrate source data. 
Chen et al. (2021)\cite{RN492} apply this approach within an empirical likelihood framework for survival analysis, using smoothly clipped absolute deviation penalties to incorporate aggregate survival constraints from source populations. 
Li et al. (2023)\cite{RN147} extend this idea to hybrid control trials with time-to-event endpoints by penalising drift between current and source controls in a Cox model using an adaptive lasso. 
Kanapka and Ivanova (2024)\cite{RN567} further extend the adaptive lasso approach to basket trials. 
Kerioui et al. (2024)\cite{RN605} propose a method for basket trials that introduces fusion penalties in logistic regression, shrinking pairwise differences between basket-specific parameters to cluster similar baskets.

\subsubsection{Combining data across populations in a weighted test statistic}

Five methods\cite{RN1544, RN1473, RN920, RN786, RN900} propose making inferences about $\theta$ using a test statistic that combines information from the source and target populations.

Let $Z_T$ and $Z_S$ denote standardised test statistics comparing mean responses in the target and source population studies, respectively. The test statistic for testing $H_0:\ \theta=0$ is defined as 
\begin{equation*}
	Z_w=\sqrt{w}Z_S+\sqrt{1-w}Z_T 
\end{equation*}
where $0 \leq w \leq 1$.

Lan et al. (2005)\cite{RN1544} propose a weighted $Z$-statistic, also considered by Huang et al. (2012)\cite{RN1473} for simultaneous global drug development programmes. 
Yu et al. (2023)\cite{RN920} extend this approach by deriving $w$ adaptively from a calibrated power prior\cite{RN418}. 
Tsou et al. (2012)\cite{RN786} propose a weighted $t$-test statistic for smaller sample sizes, with $w$ determined using Fieller's theorem. 
Gao et al. (2025)\cite{RN900} introduce an asymmetric exponential weighting scheme for constructing $Z_w$.

\subsubsection{Assessing the consistency of data before borrowing}

Seven methods\cite{RN900,RN56, RN672, RN63, RN885, RN1805} adopt a two-step strategy to assess consistency between populations before borrowing. 

A common approach is test-then-pool, which allows borrowing only when there is insufficient evidence to reject $\theta_T = \theta_S$. Liu et al. (2022)\cite{RN56} propose two methods based on propensity score matching to ensure covariate balance between source and current controls.  
Li et al. (2022)\cite{RN672} adopt a similar approach using resampling to assess outcome similarity. Okada et al. (2024)\cite{RN63} refine this by decomposing the pooling decision into two one-sided tests. 
Zhang and Li (2024)\cite{RN885} extend this to exploratory basket trials, while Gao et al. (2025)\cite{RN900} further generalise it by incorporating a weighted test statistic after the pooling decision.

Beyond test-then-pool, Zeng et al. (2020)\cite{RN1805} propose a probabilistic framework that incorporates source evidence by adaptively adjusting the significance level of the current study based on equivalence hypotheses evaluated in the source populations. 

\subsection{Causal inference methods}

{\color{black}Causal inference methods approach extrapolation as a problem of generalisability or transportability. These methods seek to account for systematic differences between populations through adjustment for observed covariates and rely on assumptions such as consistency, exchangeability, and positivity. Under these assumptions, covariate adjustment can help reduce bias arising from differences in baseline characteristics and treatment-effect modifiers between populations.} 

A total of nine causal inference methods\cite{RN426,RN221,RN13111,RN131,RN327,RN114,RN31,RN895,RN313} were identified. Of these, five\cite{RN426,RN221,RN13111,RN131,RN327} combine data across populations using causal estimators, while four\cite{RN114,RN31,RN895,RN313} propose covariate adjustment approaches to down-weight source data.  

The first group includes causal estimators, involving outcome regression, inverse probability weighting, doubly robust estimators, and augmented inverse probability weighting, which combines outcome regression and inverse probability weighting. These methods estimate treatment effects in the target population by adjusting for observed differences in baseline covariates between the source and target populations.  

The second group includes covariate adjustment approaches, involving propensity score adjustment, propensity score matching, inverse probability weighting, bootstrap-based variance adjustment, and classical covariate adjustment. Rather than directly estimating treatment effects in the target population, these approaches aim to align covariate distributions between populations before borrowing information from the source population.

\subsubsection{Combine data in a causal inference estimator}

Li et al. (2023)\cite{RN426} aim to improve efficiency and generalisability in small-sample trials by introducing a full-data doubly robust estimator under a mean exchangeability assumption for control outcomes.  
Santacatterina et al. (2025)\cite{RN221} extend this method to incorporate non-concurrent controls in platform trials. 
Zhou et al. (2024)\cite{RN13111} address open-label extension trials with control-to-treatment crossover, by proposing estimators that use propensity scores and density ratios to adaptively weight current and source data for long-term treatment effect estimation. 
Zhou et al. (2025)\cite{RN131} further develop this by proposing a two-layer weighting strategy, in which the first layer applies balancing weights to adjust for covariate differences, and the second introduces synthesis weights to combine current and source control estimates.  
Liao et al. (2025)\cite{RN327} propose a prognostic model-based approach, where source data are used to train a model that generates prognostic scores for participants in the current trial; these scores are then incorporated as covariates in an efficient estimator.

\subsubsection{Covariate adjustment for down-weighting the source datasets}

Lin et al. (2018)\cite{RN114} apply propensity score matching, after which matched source controls are combined with current controls. 
Harun et al. (2023)\cite{RN31} use a similar approach to adjust the allocation ratio during enrolment. Sachdeva et al. (2022)\cite{RN895} integrate propensity score methods with sampling-importance resampling by estimating propensity scores to balance covariates between populations and using score densities to compute importance weights for resampling source subjects before combining them with current trial data. 

Siegfried et al. (2023)\cite{RN313} propose an adjustment strategy to reduce the required sample size without directly including source patients. Their method estimates a prognostic score from source data and incorporates it as a covariate in the analysis model. The sample size is then reduced according to the variance explained by the prognostic model and the correlation between estimated and true scores.

\section{Discussion}
\label{discussion}

{\color{black}In this review, we have summarised papers that address the fundamental challenge: how to balance efficiency gains obtained through information borrowing against the risk of bias arising from differences between source and target populations \cite{RN41,gamalo2017statistical}. The key distinction between methods often lies in how source-target similarity is defined, how borrowing is controlled, and how uncertainty regarding population differences is quantified.} 

Although a wide range of methods has been proposed, no single approach is universally optimal. 
Adaptive Bayesian and frequentist methods are generally attractive when population similarity is uncertain because they can dynamically reduce borrowing under prior-data conflict. Such features are broadly consistent with emerging regulatory perspectives, which emphasise transparency of assumptions and robustness to model misspecification \cite{ICH_E11A,ICH_E11R,ema2018reflection,fda2026bayesian}. 
In contrast, non-adaptive borrowing approaches are most appropriate when there is strong clinical, biological, or pharmacological justification for assuming similarity between populations. Causal inference approaches are particularly valuable when population differences can be largely explained through observed covariates and may be especially powerful when integrated with Bayesian or frequentist borrowing frameworks. 

{\color{black}Within the Bayesian framework, borrowing is typically achieved through explicit probabilistic assumptions regarding source-target similarity. Power priors assume that source and current studies are sufficiently similar, or partially exchangeable, and regulate borrowing through a discounting parameter. Commensurate priors adopt a different perspective by explicitly modelling the relationship between source and target parameters and estimating the degree of commensurability between them. Meta-analytic predictive priors and Bayesian hierarchical models instead rely on exchangeability assumptions, whereby studies are assumed to arise from a common underlying distribution and borrowing is determined by estimated between-study heterogeneity. Although these approaches are often presented as distinct methodological classes, they can be viewed as alternative parameterisations of a common objective: controlling the extent of shrinkage between populations. For example, with a single source study, a commensurate prior can be reparameterised into a form closely related to a meta-analytic predictive prior, while both commensurate and hierarchical approaches exhibit borrowing behaviour similar to a fixed power prior when heterogeneity parameters are treated as fixed. 
These relationships illustrate that some apparently different borrowing methods may be more closely connected than suggested by their original formulations, with differences arising partly from the choice of parameterisation and modelling framework.

From a theoretical perspective, Bayesian inference may depend on the chosen parameterisation because posterior distributions are generally not invariant to transformations when non-invariant priors are used. This consideration has motivated the development of transformation-invariant priors, such as Jeffreys priors\cite{Jeffreys1946}. Consequently, although different borrowing methods may exhibit similar behaviour under certain parameterisations, their resulting inferences may not be identical when alternative prior specifications are adopted. In practice, the impact of these choices should be evaluated alongside other modelling assumptions through sensitivity analyses and robustness assessments.

Many recent developments have focused on adaptive borrowing mechanisms that attempt to address prior-data conflict and source-target heterogeneity. Robust meta-analytic predictive priors, mixture priors, Bayesian model averaging approaches, exchangeable-non-exchangeable models, and related methods all seek to reduce borrowing when differences between populations are detected. Although these approaches differ in their implementation, they share a common objective of mitigating the risk of borrowing when this is not appropriate. Such robust borrowing approaches are particularly attractive in settings where population similarity is uncertain, as they provide protection against model misspecification while retaining efficiency gains when populations are genuinely similar. 

Frequentist borrowing methods achieve similar objectives through different mechanisms. 
These methods typically rely on weighting, shrinkage, penalisation, or combined test statistics. Despite their different inferential paradigms, many frequentist borrowing procedures have close Bayesian analogues. Hence, the distinction between Bayesian and frequentist borrowing often lies less in the underlying assumptions regarding source-target similarity and more in how these assumptions are represented and how uncertainty is quantified. Bayesian approaches generally provide greater flexibility for modelling uncertainty and adaptive borrowing through prior settings, whereas frequentist approaches retain familiar inferential properties and operating characteristics. 

Causal inference methods occupy a complementary position within the extrapolation literature. Unlike traditional borrowing methods, which focus on how information is incorporated into estimation, causal inference approaches focus on whether treatment effects can be validly transported from source populations to a target population. Although many extrapolation methods were not originally developed within a causal-inference framework, concepts such as generalisability, transportability, exchangeability, positivity, and consistency provide a useful lens through which to interpret the assumptions underlying extrapolation\cite{Hernan2020,RN23}. Recent methodological developments increasingly combine causal adjustment with Bayesian or frequentist borrowing frameworks. In these approaches, source and target populations are first adjusted to achieve conditional exchangeability with respect to observed covariates, after which borrowing procedures are applied. Therefore, the borrowing assumption shifts from unconditional similarity between populations to conditional similarity given observed covariates. These methods may improve the credibility of extrapolation when important baseline differences exist between populations.

Taken together, these methodological developments have largely been motivated by specific practical challenges arising in different extrapolation settings. For example, power priors, commensurate priors, and meta-analytic predictive priors were developed to enable borrowing from source populations while accounting for uncertainty regarding source-target similarity \cite{RN430,RN311,RN325}. Robust borrowing approaches, such as mixture priors and Bayesian model averaging methods, reduce the risk of over-borrowing by incorporating uncertainty regarding the exchangeability of the source and target populations \cite{RN909,RN123}. Causal inference methods were introduced to accommodate differences in observed baseline characteristics between populations and to support the transportability of treatment effects \cite{RN218, RN908}. Similarly, methods developed for basket trials and multi-regional clinical trials were motivated by the need to borrow information across multiple related populations while allowing for potential heterogeneity \cite{RN459,RN535,RN519}. These examples show how methodological developments have often been driven by specific scientific and regulatory challenges rather than by the pursuit of a universally superior borrowing strategy.} 


{\color{black}The review also highlights several common methodological challenges that extend across statistical frameworks. First,} source-target heterogeneity and prior-data conflict remain the primary threats to valid extrapolation. Differences may arise from baseline characteristics, treatment-effect modifiers, outcome distributions, temporal changes in standard of care, or other unobserved factors. Such differences can increase bias, inflate type I error rates, and reduce the anticipated efficiency gains from borrowing. Adaptive borrowing methods attempt to address these challenges by reducing borrowing when conflict is detected, often through data-dependent weighting mechanisms. {\color{black}However, because these mechanisms use observed data to determine borrowing strength, they may themselves become sensitive to atypical target datasets. Hence, careful evaluation of operating characteristics through simulation studies and sensitivity analyses remains essential.}  

{\color{black}Second,} control of type I error represents a particularly important consideration\cite{fda2026bayesian,RN671,RN234}. It is well recognised that borrowing external information may alter type I error rates when source and target populations differ. Both Bayesian and frequentist approaches may exhibit inflated type I error under substantial prior-data conflict. Adaptive Bayesian and frequentist methods partially mitigate this risk by dynamically reducing borrowing and, in some cases, incorporating calibration procedures to improve frequentist operating characteristics. Nevertheless, the well-known robustness-efficiency trade-off implies that substantial gains in power or precision are generally difficult to achieve while simultaneously maintaining strict control of type I error under all possible violations of borrowing assumptions \cite{RN234}. 

{\color{black}At present, there is limited methodological or regulatory consensus regarding what constitutes an acceptable degree of borrowing under different contexts of use. One possible approach is to consider alternative success criteria for evaluating extrapolation methods rather than relying exclusively on strict type I error control. Recent Food and Drug Administration draft guidance on Bayesian methodology in clinical trials highlights decision-making criteria such as posterior probabilities and benefit-risk assessments based on decision-theoretic principles \cite{fda2026bayesian}. Similar considerations may also be relevant when evaluating frequentist and causal inference approaches for extrapolation, although further methodological research is needed to establish appropriate evaluation criteria across different frameworks. 

Another important consideration is the specification of the scientific question and the assumptions required to support extrapolation. In practice, extrapolation depends on the extent to which the source and target populations are believed to be sufficiently similar with respect to the estimand of interest. The acceptable degree of source-target heterogeneity, and the level of uncertainty that can be tolerated regarding population similarity, will depend on the context of use, the availability of supporting evidence, and the consequences of an incorrect inference. When substantial uncertainty remains regarding the validity of extrapolation assumptions, more robust borrowing strategies, additional sensitivity analyses, or even reliance on target-population data alone may be preferable.}



{\color{black}These considerations align closely with the recently introduced ICH M15 framework for assessing model-informed evidence\cite{ICH_M15}. 
From this perspective, extrapolation methods should not be evaluated solely according to their statistical formulation, but according to the role they play in decision-making and the potential consequences of incorrect conclusions. When only limited target-population data are available, extrapolation models may have a substantial influence on the final inference, thereby increasing model risk. Several methods provide quantitative measures of borrowing, including shrinkage weights, borrowing parameters, and effective sample size measures. For example, shrinkage estimation methods may impose lower bounds on the contribution of target-population data, thereby preventing the analysis from being completely dominated by external information\cite{RN444}. Similarly, some methods impose an upper limit on the borrowing weight to restrict the maximum contribution of external data\cite{RN397,RN918,RN217}. 
The interpretation of effective sample size measures also requires careful consideration, as different definitions quantify different aspects of borrowing. These include the prior effective sample size, which reflects the amount of information contributed by the borrowing prior; the posterior effective sample size, which reflects the effective amount of information after combining prior and target-population data; and the effective current sample size, a data-dependent measure that quantifies the impact of a prior in terms of an equivalent number of current observations \cite{RN873,RN334}. Consequently, the choice of effective sample size metric should be aligned with the specific objective of the assessment and the stage at which borrowing is being evaluated.
However, there is currently limited methodological or regulatory consensus regarding what constitutes an acceptable degree of borrowing under different contexts of use. Thus, assessment of model influence and model risk remains largely context-dependent and should consider clinical plausibility, robustness of assumptions, and regulatory objectives. 

The stage of drug development may represent an additional consideration when selecting an extrapolation strategy. In early development, where the primary objectives are often dose selection, signal detection, or proof-of-concept assessment, greater tolerance for modelling assumptions may be acceptable in exchange for increased efficiency. Consequently, model-informed approaches and methods that borrow information across endpoints or development stages may be particularly valuable. However, the present review focused on confirmatory development settings and therefore provides limited coverage of early-phase applications. Two identified papers considered borrowing information from Phase I studies, however, their primary objective was to support inference in Phase II trials \cite{RN333,RN361}; the vast majority of identified methods were developed for Phase II, Phase III, or Phase IV settings. In addition, two papers considered the use of extrapolation methods to accelerate interim analyses \cite{RN887,RN1097}. 

Despite the large number of published methods, several important methodological challenges and areas for further development remain. First, relatively little guidance exists regarding method selection. Most papers focus on proposing new methodologies and comparing them against limited benchmarks rather than conducting comprehensive head-to-head comparisons across multiple frameworks. Hence, many studies conclude that a proposed method performs well without clearly specifying the contexts in which it is preferable or the criteria by which superiority should be judged. More explicit definition of the scientific question, estimand, similarity assumptions, and performance metrics would facilitate more meaningful comparisons across methods. 
Second, there remains no consensus regarding how similarity between source and target populations should be assessed. Different methods invoke exchangeability, commensurability, transportability, covariate overlap, or prior-data compatibility, yet standardised quantitative criteria for evaluating these assumptions remain lacking. Third, although numerous adaptive borrowing procedures have been proposed, limited guidance exists regarding acceptable levels of borrowing or effective sample size contributions from external data. Fourth, most methodological evaluations continue to rely heavily on simulation studies, with relatively few prospective applications demonstrating the practical impact of extrapolation methods in regulatory decision-making. }

\section{Conclusion}
\label{conclusion}

This methodological review summarises statistical approaches for extrapolating information from a source population to support inference in a target population within confirmatory clinical trials, including settings of extrapolation across age groups or across similar indications. 
The review identified 159 papers describing 218 distinct statistical methods, which are briefly summarised in the online supplementary materials. These methods were categorised into three broad themes: Bayesian, Frequentist, and Causal Inference methods. Within each theme, the identified papers were further classified into several subgroups. Although these subgroups are, to some extent, arbitrary and the work could have been categorised in other ways, we believe this structure is useful for highlighting key methodological developments. It also provides insight into how the specific methods described in the identified papers could be generalised or applied more broadly. Furthermore, this categorisation has enabled us to identify important foundational references on which the newer work reported in detail here is based. {\color{black} A limitation of this review is that it relied exclusively on a single database, SCI-EXPANDED, which may have resulted in some relevant records being missed. As SCI-EXPANDED is a high-quality database that provides broad coverage of the relevant statistical, methodological, and biomedical literature, the impact of this limitation is likely to be modest. In addition, study screening was conducted by a single reviewer, which may have increased the risk of subjective judgement in study selection and classification.}

According to this review, no single extrapolation method is universally optimal. Method selection should be guided by the scientific question, the availability and quality of source and target data, the plausibility of similarity assumptions, the stage of development, and the intended regulatory context.
Several methodological challenges remain, including the assessment of source-target similarity, the determination of acceptable levels of borrowing, the evaluation of model risk, and the development of practical guidance for method selection. Future research should focus on establishing standardised evaluation frameworks, conducting comparative assessments across competing approaches, and generating additional evidence from prospective applications to support the reliable and transparent use of extrapolation methods in clinical development and regulatory decision-making. 

In this review, several methods represent special cases of borrowing strategies. Three\cite{RN327,RN313,RN697} use historical data to construct predictions or summary measures of the control outcome, which are then included as covariates in the analysis model for the new study. 
Basu et al. (2020)\cite{RN784} used a PK/PD model to assess the similarity between adult and paediatric populations and then constructed {\color{black}a} power prior to borrow information from the adult data. In this approach, the PK/PD model is used to quantify cross-population similarity rather than to directly perform PK/PD-based extrapolation. More generally, PK/PD-based measures of similarity could be incorporated into both Bayesian and frequentist borrowing frameworks. Such approaches may be further extended to support extrapolation in paediatric trials. 

In addition, 27 methods for basket trials are included, as these approaches can be applied to borrow from multiple source population datasets. Most of these are Bayesian adaptive down-weighting methods, with two being frequentist approaches. 
Forty-five methods use source data to inform the design and analysis of prospective trials in the target population.
Software was available for 109 of the 218 methods identified in this review, with implementations in \texttt{R}, \texttt{MATLAB}, \texttt{Python}, \texttt{WinBUGS}, \texttt{Stan}, and \texttt{SAS/STAT}.
Full details are provided in the online supplementary materials. {\color{black}Practical tutorials describing Bayesian implementation in commonly used software platforms are also available, which cover prior specification, Markov Chain Monte Carlo algorithms, and model assessment\cite{Johnston2024}.}

\bmhead{Supplementary information}

The supplementary material provides a list of the identified papers and concise descriptions of the corresponding methods.

\backmatter

\section*{Declarations}

\bmhead{Abbreviations}

\begin{enumerate}
	\item[$\bullet$]  ICH: International Council for Harmonisation of Technical Requirements for Pharmaceuticals for Human Use;
	\item[$\bullet$]  PK/PD: pharmacokinetic/pharmacodynamic; 
	\item[$\bullet$]  PRISMA: Preferred Reporting Items for Systematic reviews and Meta-Analyses;
	\item[$\bullet$]  SCI-EXPANDED: Science Citation Index Expanded.
\end{enumerate}

\bmhead{Ethics approval and consent to participate}

Not applicable.

\bmhead{Consent for publication}

Not applicable.

\bmhead{Availability of data and materials}

Not applicable.

\bmhead{Competing Interests}

FM is an employee and stockholder of F. Hoffmann-La Roche AG. AJ is an employee and stockholder of Novartis AG (Basel). XZ is an employee and stockholder of Novartis AG (East Hanover). MV is an employee of Bayer BCC AG. NS acknowledges research support received in the past three years from NIHR, MRC, and the UKRI Horizon Europe Guarantee scheme. LW, CR, MP, and SZ declare no competing interests. 

\bmhead{Funding}

This work was supported by the INVENTS project, which has received funding from the European Union's Horizon Europe Research and Innovation Programme. Additional funding was provided by the Swiss State Secretariat for Education, Research and Innovation (SERI) and by UK Research and Innovation (UKRI) under the Horizon Europe Guarantee scheme.

\bmhead{Authors' Contributions}

LW and NS constructed the identification and screening for the methodological review. LW drafted the original manuscript, while NS reviewed, edited, and supervised the work. FM, AJ, XZ, CR, MV, MP, and SZ provided comments on an earlier version of the manuscript. All authors contributed to manuscript revision, read, and approved the final manuscript.

\bmhead{Clinical trial number}
Not applicable.










\begin{appendices}
	
	\section{Web of Science search terms used in literature search}\label{search terms}
	
	\begin{enumerate}
		\item[\bf Search 1:] (TS=( (``historical control*" OR ``historical information" OR ``historical data" OR ``borrow*" OR ``extrapolat*") AND (trial*) )) AND (WC=(Biology OR Mathematical \& Computational Biology OR Mathematics, Applied OR Mathematics, Interdisciplinary Applications OR Medical Informatics OR Medicine, Research \& Experimental OR Statistics \& Probability))
		
		\item[\bf Search 2:] (TS=( (bridging OR ``borrow* strength" OR extrapolat* OR synthesize) AND (p\$ediatric OR child* OR ethnic OR region* OR geotherapeutic* OR centre OR center) AND (trial* OR ``bridging stud*") )) AND (WC=(Biology OR Mathematical \& Computational Biology OR Mathematics, Applied OR Mathematics, Interdisciplinary Applications OR Medical Informatics OR Medicine, Research \& Experimental OR Pediatrics OR Statistics \& Probability))
		
		\item[\bf Search 3:] (TS=( (short-term endpoint OR short-term end point OR biomarker OR surrogate endpoint OR surrogate end point) AND (long-term endpoint OR long-term end point) AND (trial*) )) AND (WC=(Biology OR Mathematical \& Computational Biology OR Mathematics, Applied OR Mathematics, Interdisciplinary Applications OR Medical Informatics OR Medicine, Research \& Experimental OR Pediatrics OR Statistics \& Probability)) 
	\end{enumerate}
	
	
	
	
\end{appendices}

\end{document}